# BiEntropy – The Approximate Entropy of a Finite Binary String

v2.33

Grenville J. Croll

grenvillecroll@gmail.com

*We design, implement and test a simple algorithm which computes the approximate entropy of a finite binary string of arbitrary length. The algorithm uses a weighted average of the Shannon Entropies of the string and all but the last binary derivative of the string. We successfully test the algorithm in the fields of Prime Number Theory (where we prove explicitly that the sequence of prime numbers is not periodic), Human Vision, Cryptography, Random Number Generation and Quantitative Finance.*

## 1 INTRODUCTION

The purpose of this paper is to illustrate the means by which a finite binary string of arbitrary length can be compared against another in terms of the relative order and disorder of all of its digits. We do this using a simple function called BiEntropy, which is based upon a weighted average of the Shannon Entropies of all but the last binary derivative of the string.

This paper is organised as follows: First, we briefly cover the historical background regarding the development of tests and measures of order, disorder, randomness, irregularity and entropy. We show that binary derivatives have previously been used in the measurement of disorder and have also been used in cryptographic applications and attacks.

Second, we discuss our intuitive understanding of order and disorder, Shannon Entropy and Binary Derivatives in more detail. We outline a number of issues that were important in formulating the BiEntropy function, following which we formally define it together with some simple variations.

Third, we apply the BiEntropy method in a number of diverse application areas including:

a) *Prime Number Theory* – We show that BiEntropy empirically determines that the sequence of early prime numbers is not periodic. This result is implicit due to the Prime Number Theorem. We prove this non-periodicity explicitly through two simple corollaries.

b) *Human Vision* - We demonstrate the use of BiEntropy in powerfully discriminating between the geometrical layouts of some standard ISO dot matrix characters compared to some randomly produced strings, the Braille character set and two specially designed high and low entropy character sets.

c) *Random Number Generation* - We use BiEntropy to evaluate the decimal expansions of some well known random, normal and irrational numbers using a simple decimal to binary coding scheme. We compare these expansions with similar output from the Random Number Generators within Excel 2003 and Excel 2010.

d) *Cryptography* - We use BiEntropy to simply and easily reveal significant differences between the encrypted and unencrypted binary files of some real and synthetic spreadsheets.

e) *Quantitative Finance* - We use BiEntropy to examine 10 years worth of UK daily stock market prices to show that the BiEntropy of historical price changes is strongly correlated with future stock market prices.

Finally, we provide a summary and identify some areas for future work.

## 2 HISTORICAL BACKGROUND

There are very many tests and algorithms for the determination or measurement of randomness, regularity, irregularity, order, disorder and entropy for binary and other strings [Marsaglia, 1968][Gao, Kontoyiannis & Bienenstock, 2008]. There are several measures which derive a scalar linked to the randomness, disorder or entropy of finite strings such as Binary Entropy [Shannon, 1948], Approximate Entropy [Pincus, 1991] [Pincus & Singer, 1996][Rukhin, 2000a,b], Sample Entropy [Richman &Moorman, 2000] and Fuzzy Entropy [Chen, Wu & Yang, 2009]. Existing tests and algorithms appear to be divided into two classes: those which use a sliding window technique to examine substrings of the original string [Marsaglia & Zaman, 1993] and those which are related to the length of algorithms used to generate the entire string [Kolmogorov, 1965] [Chaitin, 1966]. Some of these functions have been applied in critically important domains [Pincus & Viscarello, 1992].

Whilst there have been a number of attempts to use binary derivatives in randomness tests [McNair, 1989], cryptographic applications [Carroll, 1989, 1998] and attacks [Bruwer, 1995], we believe that the use of a weighted average of the Shannon entropies of the binary derivatives of a string is unique.

## 3 INTUITIVE INSIGHT INTO BINARY ORDER & DISORDER

Table 1 suggests how we might intuitively regard the order and disorder of some 8 bit binary strings.

**Table 1 - Intuitive Insight into some short binary strings**

| Binary String | Description | Reason |
|---|---|---|
| 11111111 | Perfectly ordered | All 1's |
| 00000000 | Perfectly ordered | All 0's |
| 01010110 | Mostly ordered | Mostly 01's |
| 01010101 | Regular, not disordered | Repeating 01's |
| 11001100 | Regular, not disordered | Repeating 1100's |
| 01011010 | Mostly ordered | 0101 then 1010 |
| 01101011 | Somewhat disordered | No Apparent Pattern |
| 10110101 | Somewhat disordered | No Apparent Pattern |

Determination of the relative order and disorder of the 256 possible 8-bit binary strings is an obvious example problem that does not appear to have been previously addressed in the literature. There are 256! (≈8.58 * $10^{508}$) differing ways of ordering the 8 bit binary strings.

We require an algorithm which will determine the relative and/or absolute degree of order and disorder of binary strings such as the above, for arbitrarily long binary strings. The algorithm will return 0 for perfectly ordered strings and 1 for perfectly disordered strings.

## 4 SHANNON ENTROPY, BINARY DERIVATIVES & WEIGHTING METHODS

### 4.1 Shannon Entropy

Shannon's Entropy of a binary string $s = s_1, . . . ,s_n$ where $P(s_i=1) = p$ (and $0 \log_2 0$ is defined to be 0) is:

$$H(p) = -p \log_2 p - (1-p) \log_2 (1-p)$$

For perfectly ordered strings which are all 1's or all 0's i.e. $p = 0$ or $p = 1$, $H(p)$ returns 0. Where $p = 0.5$, $H(p)$ returns 1, reflecting maximum variety. However, for a string such as 01010101, where $p = 0.5$, $H(p)$ also returns 1, ignoring completely the periodic nature of the string.

## 4.2 Binary Derivatives & Periodicity

The first binary derivative of $s$, $d_1(s)$, is the binary string of length $n - 1$ formed by XORing adjacent pairs of digits. We refer to the $k$th derivative of $s$ $d_k(s)$ as the binary derivative of $d_{k-1}(s)$. There are $n$-1 binary derivatives of $s$.

Some years ago [Nathanson, 1971], following the work of [Goka, 1970] defined the notions of *period* and *eventual period* within arbitrary binary strings and outlined the related properties of the derivatives both individually and collectively. Amongst a number of useful results we find that: a) if the derivative of a binary string is eventually periodic with a period $P$ then the binary string is also eventually periodic with a period $P$ or $2P$; b) if a derivative is all zero's then the string has a period $2^m$ for some $m$, $0 \leq m \leq n$; c) if a derivative has eventual period $P$, the string has eventual period $2^m P$ for some $m$ satisfying $0 \leq m \leq n$.

Adapting Nathanson's definitions for finite strings, a binary string $s$ of length $n$ is *periodic* if, for some least positive integer $P$, $s_{i+P} = s_i$ for all $1 \leq i \leq n-P$. A binary string $s$ of length $n$ is *eventually periodic* if, for some least positive integer $P$ and some least positive nonnegative integer $k$, $s_{i+P} = s_i$ for all $k \leq i \leq n-P$. Note that a finite binary string can be read left to right or right to left such that we *may* need to refer to a string as being either *right* or *left reading eventually periodic* and adapt our notation accordingly.

For example, the first binary derivative of 01010101 (with *period*, $P = 2$) is 1111111 ($P = 1$), following which all the higher derivatives are all 0's. The third derivative of 00010001 ($P = 4$) is 11111, following which again all the higher derivatives are 0. The sixth derivative of 00011111 (with *right reading eventual period P* = 1 from the fourth digit) is 10.

By calculating all the binary derivatives of $s$ we can discover the existence of repetitive patterns in binary strings of arbitrary length. If a binary string is *periodic* its last derivative is zero. A binary string is *aperiodic* if its last derivative is 1 (else its last but one derivative is periodic and the string itself is therefore eventually periodic). A binary string is *nperiodic* if its last derivative is 0, but is not periodic.

Although Nathanson's definitions (and our own adapted definitions) of periodicity and eventual periodicity are useful, in this paper we rely solely upon the binary derivatives of a finite string to resolve the issue of the periodicity within the string.

[Davies et al, 1995] outline some further properties of binary derivatives. Let $p(k)$ denote the observed fraction of 1's in $d_k(s)$ where $p(0)$ denotes the fraction of 1's in $s$. Let $\pi(k)$ denote the corresponding population proportions. Provided $p(0) = 0.5$, $p(1)$ is not correlated with $p(0)$. Likewise, where $\pi(k) = 0.5$, $p(k+1)$ is not correlated with $p(k)$, the sample proportions are independent. By induction, these properties apply to the higher derivatives.

## 4.3 Weighting Methods

Thus there are number of important factors to consider in designing a function $\mathcal{F}$ to compute the approximate entropy of a finite binary string:

i) The Shannon entropy of a binary string does not give a complete picture regarding the order and disorder of the string due to the failure to accommodate periodicity.
ii) The derivatives of a binary string determine the existence of periodicity in the string.
iii) Determination of any periodicity may require the evaluation of all $n$-1 derivatives.
iv) The proportion of 1's & 0's in the binary derivatives are or can be independent.
v) Differing consideration may have to be given to higher or lower derivatives
vi) The function is required to be effective on strings of arbitrary length.

The Shannon entropy of a binary derivative is H($p(k)$). The approximate entropy of a binary string – the BiEntropy – could therefore be some function $\mathcal{F}$(H($p(k)$)) for $0 \leq k < n$.

We must decide how to combine or weight the H($p(k)$) in order to arrive at a function $\mathcal{F}$ that is likely to have some utility. The field of time series analysis [Makridakis et al, 2008] provides comprehensive guidance on a variety of methods (including moving averages & exponential smoothing) which are used to extract information from sometimes noisy historical time series.

The H($p(k)$) of a string for increasing or decreasing $k$ is clearly a progression, possibly noisy, though not temporal. Exponential weighting would therefore be a first choice of weighting method. Exponential methods have the advantage that they can accommodate numerical series of arbitrary length – no matter how long the string all the H($p(k)$) would make some contribution.

A competing consideration is that each of the H($p(k)$) is potentially independent of any other and so any weights must discriminate clearly between each H($p(k)$). Although we could manipulate an exponential method to do this with an additional parameter, a simpler method would be to assign a polynomial weight such as $2^k$ to each of the H($p(k)$) (which vary between 0 and 1) thereby clearly separating the influence of each from the other.

Since the H($p(k)$) are not temporal, we could weight the H($p(k)$) from the highest ($k = n$-1) to the lowest ($k = 0$) derivative or vice versa. For higher periods $P$, the $d_k$ only fall to zero at a higher $k$. For some strings the $k$-1$^{th}$ derivative does not fall to zero at all, indicating that there is no periodicity in the binary string. Since we are attempting to measure the order and disorder of a binary string, if no order (i.e. periodicity) has emerged following the calculation of the $n$-1$^{th}$ derivative we should assign the highest weight to that derivative thereby indicating that that string is more disordered than other strings where the derivative falls to zero earlier (i.e. at a lower $k$) and is either periodic or eventually periodic.

Other weighting methods could include *none* where the H($p(k)$) are simply averaged and *linear*, where the H($p(k)$) is a constant proportion of $k$. We have not evaluated either of these latter two methods.

## 5 BIENTROPY

BiEntropy, or BiEn for short, is a weighted average of the Shannon binary entropies of the string and the first $n$-2 binary derivatives of the string using a simple power law. This version of BiEntropy is suitable for shorter binary strings where $n \leq 32$ approximately.

$$\text{BiEn}(s) = \left(1 / (2^{n-1} - 1)\right)\left(\sum_{k=0}^{n-2} ((-p(k) \log_2 p(k) - (1-p(k)) \log_2 (1-p(k)))) 2^k\right)$$

The final derivative $d_{n-1}$ is not used as there is no variation in the contribution to the total entropy in either of its two binary states. The highest weight is assigned to the highest derivative $d_{n-2}$.

If the higher derivatives of an arbitrarily long binary string are periodic, then the whole sequence exhibits periodicity. For strings where the latter derivatives are not periodic, or for all strings in any case, we can use a second version of BiEntropy, which uses a Logarithmic weighting, to evaluate the complete set of a long series of binary derivatives.

$$\text{Tres BiEn}(s) = \left(1 / \sum_{k=0}^{n-2} \log_2 (k+2)\right)\left(\sum_{k=0}^{n-2} (-p(k) \log_2 p(k) - (1-p(k)) \log_2 (1-p(k))) \log_2 (k+2)\right)$$

The logarithmic weighting or (TBiEn for short) again gives greater weight to the higher derivatives. Depending upon the application, other weightings could be used.

The BiEntropy algorithm evaluates the order and disorder of a binary string of length $n$ in O($n^2$) time using O($n$) memory.

## 6 BIENTROPY OF THE 2-BIT STRINGS

The BiEntropy of a 2-bit string is given in Table 2.

Table 2 - The BiEntropy of a 2-bit string

| String | Description | BiEntropy |
|---|---|---|
| 00 | Perfectly ordered | 0 |
| 01 | Perfectly disordered | 1 |
| 10 | Perfectly disordered | 1 |
| 11 | Perfectly ordered | 0 |

Table 2 depicts the XOR operation and the computation of the binary derivative of a 2-bit string.

## 7 BIENTROPY OF THE 4-BIT STRINGS

We show in Tables 3A & 3B the layout of some simple Excel spreadsheets to compute the BiEn and TBiEn of a 4-bit string. We used a simple =IF statement to compute each bit of the derivatives. We show a graphic of the BiEn of the 4-bit strings in Figure 1. The TBiEn graphic is very similar though the values of BiEn and TBiEn differ slightly.

Table 3A - Computing the BiEn of a 4-bit string

| | | | | 1's | $n$ | $p$ | $(1-p)$ | $-p\log(p)$ | $-(1-p)\log(1-p)$ | BiEn | $k$ | $2^k$ | BiEn*$2^k$ |
|---|---|---|---|---|---|---|---|---|---|---|---|---|---|
| 1 | 0 | 1 | 1 | 3 | 4 | 0.75 | 0.25 | 0.31 | | 0.50 | 0.81 | 0 | 1 | 0.81 |
| 1 | 1 | 0 | | 2 | 3 | 0.67 | 0.33 | 0.39 | | 0.53 | 0.92 | 1 | 2 | 1.84 |
| 0 | 1 | | | 1 | 2 | 0.50 | 0.50 | 0.50 | | 0.50 | 1.00 | 2 | 4 | 4.00 |
| | | | | | | | | | | | 7.00 | | | 6.65 |
| | | | | | | | | | | BiEn($s$) | | | | **0.95** |

Table 3B - Computing the TBiEn of a 4-bit string

| | | | | 1's | $n$ | $p$ | $(1-p)$ | $-p\log(p)$ | $-(1-p)\log(1-p)$ | BiEn | $k$ | $\log(k+2)$ | BiEn*$\log(k+2)$ |
|---|---|---|---|---|---|---|---|---|---|---|---|---|---|
| 1 | 0 | 0 | 1 | 2 | 4 | 0.50 | 0.50 | 0.50 | | 0.50 | 1.00 | 0 | 1.00 | 1.00 |
| 1 | 0 | 1 | | 2 | 3 | 0.67 | 0.33 | 0.39 | | 0.53 | 0.92 | 1 | 1.58 | 1.46 |
| 1 | 1 | | | 2 | 2 | 1.00 | 0.00 | 0.00 | | 0.00 | 0.00 | 2 | 2.00 | 0.00 |
| | | | | | | | | | | | 4.58 | | | 2.46 |
| | | | | | | | | | | TBiEn($s$) | | | | **0.54** |

**Figure 1 - BiEn of the four-bit strings**

| BiEn of an 4-bit sequence | | BiEn2 | 0.00 | 0.00 | 1.00 | 1.00 |
|---|---|---|---|---|---|---|
| | | N | 0 | 3 | 1 | 2 |
| | | 2 | 0 | 1 | 0 | 1 |
| | | 1 | 0 | 1 | 1 | 0 |
| BiEn2 | N | 2 1 | | | | |
| 0.00 | 0 | 0 0 | 0 | 0.41 | 0.95 | 0.95 |
| 0.00 | 3 | 1 1 | 0.41 | 0 | 0.95 | 0.95 |
| 1.00 | 1 | 0 1 | 0.95 | 0.95 | 0.14 | 0.41 |
| 1.00 | 2 | 1 0 | 0.95 | 0.95 | 0.41 | 0.14 |

There are two perfectly ordered strings 0000 & 1111, two nearly ordered, periodic strings 0101 and 1010, four intermediately disordered (nperiodic) strings where the left two bits are the 1's complement of the right two bits and eight disordered (aperiodic) strings where either a single 1 or a single 0 transits a four bit field. Note the general XOR structure of the table. Mean BiEn for the 4-Bit strings is 0.594, standard deviation 0.389. Mean TBiEn is 0.644, standard deviation 0.355.

## 8 BIENTROPY OF THE 8-BIT STRINGS

We show in Figure 2 the BiEn of all 256 8-Bit strings. They are colour coded such that the *periodic* strings are white, *nperiodic* strings are light and dark yellow and *aperiodic* strings are dark orange. The diagram is structured such that the X and Y axes show the 4 bit strings of which each 8 bit string is comprised. The X and Y axes are sorted so that low BiEn or ordered 4 bit strings appear towards the top and left of the table and high BiEn or disordered 4-bit strings appear to the bottom and right. The Y-Axis corresponds to the first four bits of the string. Note that Figure 2 also has the general configuration of the XOR function. The white diagonal shows the zero and lower BiEn of the 16 repeated 4-bit strings. The TBiEn diagram is identical in its main XOR partitions but differs slightly in the other two partitions.

**Figure 2 BiEn of the 4 and 8-bit strings**

| | | BiEn4 | 0 | 0 | 0.14 | 0.14 | 0.41 | 0.41 | 0.41 | 0.41 | 0.95 | 0.95 | 0.95 | 0.95 | 0.95 | 0.95 | 0.95 | 0.95 |
|---|---|---|---|---|---|---|---|---|---|---|---|---|---|---|---|---|---|---|
| | | N | 0 | 15 | 5 | 10 | 3 | 6 | 9 | 12 | 1 | 2 | 4 | 7 | 8 | 11 | 13 | 14 |
| BiEn of an 8-bit sequence | | 8 | 0 | 1 | 0 | 1 | 0 | 0 | 1 | 1 | 0 | 0 | 0 | 0 | 1 | 1 | 1 | 1 |
| | | 4 | 0 | 1 | 1 | 0 | 0 | 1 | 0 | 1 | 0 | 0 | 1 | 1 | 0 | 0 | 1 | 1 |
| | | 2 | 0 | 1 | 0 | 1 | 1 | 1 | 0 | 0 | 0 | 1 | 0 | 1 | 0 | 1 | 0 | 1 |
| | | 1 | 0 | 1 | 1 | 0 | 1 | 0 | 1 | 0 | 1 | 0 | 0 | 1 | 0 | 1 | 1 | 0 |
| BiEn4 | | 8 4 2 1 | | | | | | | | | | | | | | | | |
| 0.00 | 0 | 0 0 0 0 | 0 | 0.11 | 0.23 | 0.23 | 0.45 | 0.45 | 0.47 | 0.45 | 0.92 | 0.94 | 0.95 | 0.94 | 0.93 | 0.95 | 0.95 | 0.95 |
| 0.00 | 15 | 1 1 1 1 | 0.11 | 0 | 0.23 | 0.23 | 0.45 | 0.47 | 0.45 | 0.45 | 0.95 | 0.95 | 0.95 | 0.93 | 0.94 | 0.95 | 0.94 | 0.92 |
| 0.14 | 5 | 0 1 0 1 | 0.23 | 0.23 | 0.01 | 0.11 | 0.46 | 0.45 | 0.45 | 0.47 | 0.95 | 0.94 | 0.92 | 0.94 | 0.95 | 0.95 | 0.93 | 0.95 |
| 0.14 | 10 | 1 0 1 0 | 0.23 | 0.23 | 0.11 | 0.01 | 0.47 | 0.45 | 0.45 | 0.46 | 0.95 | 0.93 | 0.95 | 0.95 | 0.94 | 0.92 | 0.94 | 0.95 |
| 0.41 | 3 | 0 0 1 1 | 0.45 | 0.45 | 0.46 | 0.47 | 0.02 | 0.23 | 0.23 | 0.11 | 0.94 | 0.93 | 0.95 | 0.95 | 0.95 | 0.93 | 0.95 | 0.95 |
| 0.41 | 6 | 0 1 1 0 | 0.45 | 0.47 | 0.45 | 0.45 | 0.23 | 0.02 | 0.11 | 0.23 | 0.95 | 0.95 | 0.94 | 0.93 | 0.95 | 0.95 | 0.95 | 0.93 |
| 0.41 | 9 | 1 0 0 1 | 0.47 | 0.45 | 0.45 | 0.45 | 0.23 | 0.11 | 0.02 | 0.23 | 0.93 | 0.95 | 0.95 | 0.95 | 0.93 | 0.94 | 0.95 | 0.95 |
| 0.41 | 12 | 1 1 0 0 | 0.45 | 0.45 | 0.47 | 0.46 | 0.11 | 0.23 | 0.23 | 0.02 | 0.95 | 0.95 | 0.93 | 0.95 | 0.95 | 0.95 | 0.93 | 0.94 |
| 0.95 | 1 | 0 0 0 1 | 0.93 | 0.94 | 0.94 | 0.95 | 0.95 | 0.95 | 0.93 | 0.95 | 0.05 | 0.46 | 0.24 | 0.46 | 0.47 | 0.24 | 0.46 | 0.11 |
| 0.95 | 2 | 0 0 1 0 | 0.95 | 0.95 | 0.95 | 0.92 | 0.93 | 0.94 | 0.95 | 0.95 | 0.46 | 0.05 | 0.45 | 0.24 | 0.24 | 0.47 | 0.11 | 0.46 |
| 0.95 | 4 | 0 1 0 0 | 0.94 | 0.95 | 0.93 | 0.94 | 0.95 | 0.95 | 0.95 | 0.93 | 0.24 | 0.45 | 0.05 | 0.46 | 0.46 | 0.11 | 0.47 | 0.24 |
| 0.95 | 7 | 0 1 1 1 | 0.95 | 0.92 | 0.95 | 0.95 | 0.94 | 0.95 | 0.95 | 0.95 | 0.46 | 0.24 | 0.46 | 0.05 | 0.11 | 0.46 | 0.24 | 0.47 |
| 0.95 | 8 | 1 0 0 0 | 0.92 | 0.95 | 0.95 | 0.95 | 0.95 | 0.95 | 0.93 | 0.94 | 0.47 | 0.24 | 0.46 | 0.11 | 0.05 | 0.46 | 0.24 | 0.46 |
| 0.95 | 11 | 1 0 1 1 | 0.95 | 0.94 | 0.94 | 0.93 | 0.95 | 0.95 | 0.95 | 0.95 | 0.24 | 0.47 | 0.11 | 0.46 | 0.46 | 0.05 | 0.45 | 0.24 |
| 0.95 | 13 | 1 1 0 1 | 0.95 | 0.95 | 0.92 | 0.95 | 0.95 | 0.95 | 0.94 | 0.93 | 0.46 | 0.11 | 0.47 | 0.24 | 0.24 | 0.45 | 0.05 | 0.46 |
| 0.95 | 14 | 1 1 1 0 | 0.94 | 0.93 | 0.95 | 0.94 | 0.95 | 0.93 | 0.95 | 0.95 | 0.11 | 0.46 | 0.24 | 0.47 | 0.46 | 0.24 | 0.46 | 0.05 |

Using the BiEn (and TBiEn) metrics, exactly half of the 8-bit strings are classified as being nearly perfectly disordered (BiEn > 0.90). The last binary derivative of each of these strings is 1. They are the *aperiodic* 8 bit strings. 16 strings, which comprise two repetitions of the same 4-bit string, are nearly perfectly ordered (BiEn < 0.10). The last derivative of each of these strings is 0. These are the *periodic* strings. The remainder are neither ordered nor disordered to a greater or lesser degree. These are the *nperiodic* strings – the last derivative is 0 but the entire string has no single period. BiEntropy is fractal from the self-similarity exhibited in Figures 1 & 2. Mean BiEn for the 8-Bit strings is 0.625, standard deviation 0.340. Mean TBiEn is 0.747, standard deviation 0.209. The BiEn and TBiEn of the 8 bit strings are strongly correlated (Adjusted $R^2$ = 0.85).

We show in Figure 3 the distribution of BiEn and TBien for the 8-bit strings. BiEn & TBiEn do not reach 1.0. Were they to do so, the $p(k)$ for all $k \leq n$-2 would have to be exactly 0.5, which is impossible as $k$ is odd at least once for all $n \geq 3$. Note that in the absence of a closed form solution for determining the BiEntropy of a finite binary string, BiEntropy has to be determined empirically.

**Figure 3 -  Distribution of the BiEntropy of the 8-bit Strings**

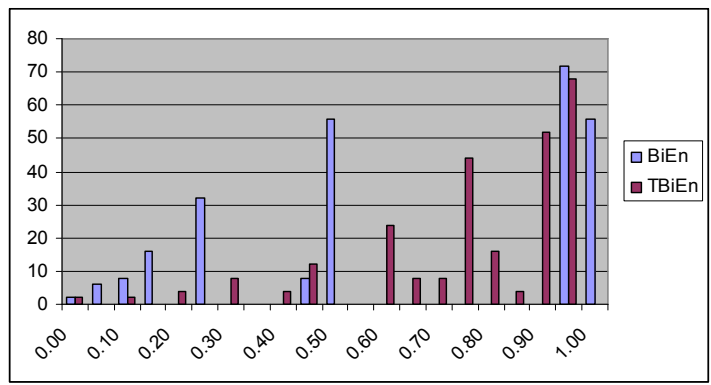

Figure 4 shows the BiEntropies of the 8-bit strings in ascending order and further illustrates the fractal nature of BiEntropy.

**Figure 4 BiEntropy of  the 8-bit strings in ascending order**

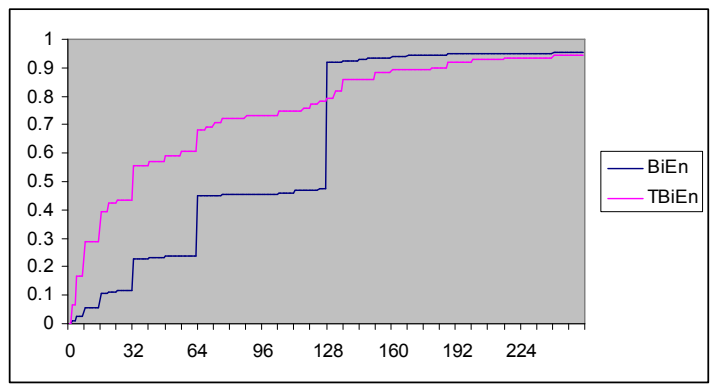

BiEn and TBiEn differ somewhat in their distributions and the way in which they order the nperiodic 8 bit strings. They order the periodic and aperiodic 8 bit strings in the same way. The comparative utility of BiEn and TBiEn will have to be determined experimentally.

## 9 BIENTROPY OF THE PRIME NUMBER SEQUENCE

### 9.1 The Binary Encoded Primes (BEP's)

Consider the $q$ natural numbers starting from 2:

2, 3, 4, 5, 6, 7, 8, 9, 10, 11, 12, 13, 14, 15, 16, 17, 18, 19…..

We can encode them in a binary string $B$ of length $n$ ($1 \leq n \leq q$) such that the primes are encoded as 1 and the composites as 0. $B_i$ is the $i$th digit of $B$ ($i \geq 1$).

1, 1, 0, 1, 0, 1, 0, 0, 0, 1, 0, 1, 0, 0, 0, 1, 0, 1…..

Thus $B_1$, corresponding to the natural number 2 is 1 and $B_4$, corresponding to the natural number 5 is also 1. We can easily compute the logarithmic BiEntropy of strings $B$ for all $n$ ($2 \leq n \leq q$) which we show in Figure 2 for $q = 512$.

**Figure 5 – The Logarithmic BiEntropy (TBiEn) of binary encoded primes < 512**

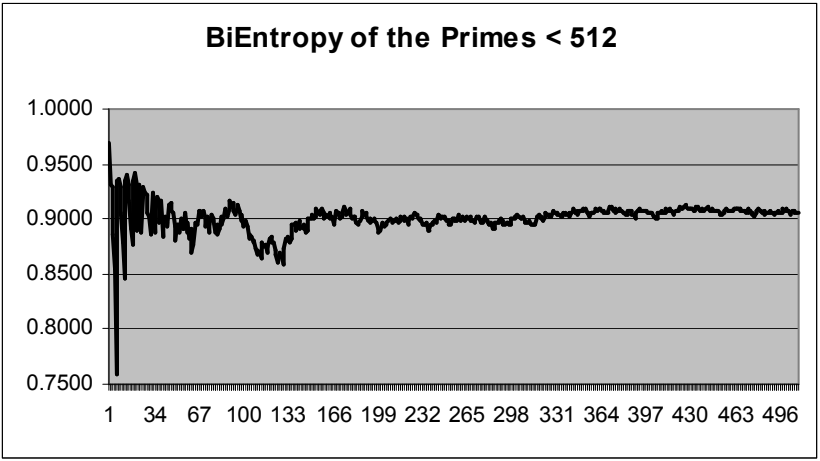

We can see that the logarithmic BiEntropies of the binary strings corresponding to the primes < 512 are close to 1.0 and that these strings are mostly (but not exclusively) aperiodic. This result is implicit due to the Prime Number Theorem. The BiEntropy dip around 114-136 corresponds to a long sequence of composites broken by only two primes.

There are some simple corollaries regarding the periodicity of the primes which follow directly from Nathanson's definitions of periodicity for infinite binary strings.

**COROLLARY ONE – THE SEQUENCE OF PRIME NUMBERS IS NOT PERIODIC**

Consider a binary string $B$ of even length $n$ ($n \geq 4$) containing the binary encoding of the primes as above starting from 2. $B_i$ is the $i$th digit of $B$ ($i \geq 1$). The binary string $B$ is periodic if, for some positive integer $p$, $B_{i+p} = B_i$ for all $1 \leq i \leq n$.

This is impossible for $p = 1$ because the even numbers are composite. This is also impossible for $p \geq 2$ because:

  a) both $B_1 = 1$ and $B_2 = 1$ (because both 2 and 3 are prime) *and*
  b) no further pairs of natural numbers which are adjacent primes can occur because the even numbers are composite.

Hence the binary string *B* corresponding to the primes is not periodic for all even values of *n* (*n* ≥ *4*) for all *p* ≥ 2. The binary string *B* of length *n* = 2 corresponding to the first two primes 2 & 3 is periodic with *p* = 1. We choose not to compare strings of unequal length which would be necessary if *n* was odd.

**9.2 The Prime Encoded Non-Negative Integers (PENNI's)**

Consider the *r* non negative integers:

0, 1, 2, 3, 4, 5, 6, 7, 8, 9, 10, 11, 12, 13, 14, 15…..

We can encode them in a binary string *E* (after Eratosthenes) of length *n* (1 ≤ *n* ≤ *r*) such that the primes are encoded as 1 and the non primes (0 and 1) and composites are encoded as 0. $E_j$ is the *j*th digit of *E* (*j* ≥ 1).

0, 0, 1, 1, 0, 1, 0, 1, 0, 0, 0, 1, 0, 1, 0, 0…..

Thus $E_1$, corresponding to 0 is 0 and $E_4$, corresponding to the natural number 3 is 1. We can again compute the logarithmic BiEntropy of strings *E* for all *n* (2 ≤ *n* ≤ 512) with results almost identical to those depicted in Figure 2 and omitted for brevity.

There follows a second simple, but differing corollary, which demonstrates that the PENNI's (which includes the primes) are not periodic, but in a way that emphasises the complete absence of all periodicity.

**COROLLARY TWO – THE PENNI'S ARE NOT PERIODIC**

Consider a binary string *E* of even length *n* (*n* ≥ 4) containing the binary encoding of the primes and the non-negative integers as above starting from 0. $E_j$ is the *j*th digit of *E* (*j* ≥1). The binary string *E* is periodic if, for some positive integer *p*, $E_{j+p} = E_j$ for all 1 ≤ *j* ≤ *n*. Let $e_{p,k,j}$ be the *j*th character of the *k*th period of *E* for period of length *p* (*p, k , j* ≥ 1). We show in Figure 6 the periodicities of the primes and the composites for various small *p* starting from the origin, 0. There is no other natural number which is the origin of all the primes and all of the composites. We choose not to compare strings of unequal length which would be necessary if *n* was odd.

**Figure 6 The Periodicities of the Primes and the Composites**

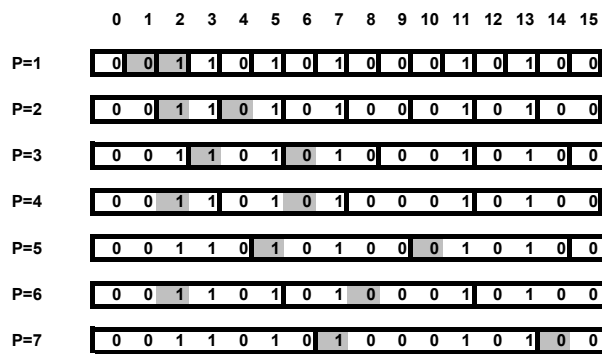

Note that for each period *p* where *p* is prime and *p* ≥ 2, $e_{p,2,1}$ is not equal to $e_{p,3,1}$ because $e_{p,2,1}$ marks the first occasion where *p* is known to be prime, after the algorithm of Eratosthenes, and $e_{p,3,1}$ marks the first non prime multiple of *p* which occurs in the first occasion of the following period. Where *p* is even $e_{p,1,3}$ is not equal to $e_{p,2,3}$ because 2 is the only even prime. For *p* = 1, there is an absence of periodicity not least because both the even numbers and the odd numbers are periodic with period 2.

Hence the binary string *E* corresponding to the PENNI's is not periodic for all even *n* ( $n \geq 4$ ) for all *p* $\geq 1$. Corollary One proves explicitly that the sequence of prime numbers is not periodic. Corollary Two is necessary to demonstrate explicitly the total absence of periodicity in the sequence of prime numbers for all even sequences for all periods from the origin 0.

## 10 BIENTROPY OF SOME 7*5 DOT MATRIX PRINTER CHARACTERS

We obtained [Mitchell, 2008] the binary patterns corresponding to the 96 Alphabetic (upper and lower case), Numeric and Punctuation characters of the US ISO 646 7-bit character set. We arranged each of these characters in a 7*5 array and computed the Horizontal and Vertical Binary BiEntropies of these strings using Horizontal and Vertical Raster scans, both of which were 35 bit binary strings.

By way of comparison, we used the random number generator within Microsoft Excel 2003 to create a set (RANDOM) of 96 randomly generated 7*5 dot matrix characters where *p*(0) of the 35 bit array was 0.5. We also used the 6 bit Braille [Jiménez et al, 2009] dot-pattern (BRAILLE), arranged as 3*2 array superimposed on the central 5*3 bits of a 7*5 grid. Finally, we designed twelve 7*5 dot matrix characters to exhibit High Entropy (HECS) and a further twelve to exhibit Low Entropy (LECS). We show samples from each character set in Table 4A & their average BiEntropies in Table 4B.

**Table 4A – Samples from five character sets with differing BiEntropies**

**Table 4B - The BiEntropy of Some 7*5 Dot Matrix Character Sets**

| Character Set | BiEn Mean | BiEn Stdev | TBiEn Mean | TBiEn Stdev | N |
|---|---|---|---|---|---|
| HECS | 0.945 | 0.006 | 0.947 | 0.011 | 12 |
| RANDOM | 0.634 | 0.240 | 0.876 | 0.168 | 96 |
| ISO | 0.494 | 0.291 | 0.844 | 0.117 | 96 |
| BRAILLE | 0.068 | 0.053 | 0.661 | 0.124 | 64 |
| LECS | 0.014 | 0.012 | 0.622 | 0.205 | 12 |

The BiEntropies of each of the five character sets are distinguished from each other (*p* < 0.01) by the BiEn metric, though note that HECS and LECS were designed to be such. For the TBiEn metric, HECS & ISO, ISO & BRAILLE and ISO & LECS were distinguished from each other (*p* < 0.01). BiEn penalises zero upper derivatives more heavily than TBiEn but note that both metrics place the groups in the same order. Note that binary strings with a length ≥ 35 may contain these characters during the evaluation of their 35$^{th}$ last derivative.

## 11 BIENTROPY OF SOME IRRATIONAL, NORMAL AND PSEUDORANDOM NUMBERS

We obtained the first million digits of the decimal expansions of π, e, √2 and √3 from the internet [Nemiroff & Bonnell, 1994][Andersson, 2013]. We also obtained a set of one million random decimal digits produced by the RAND corporation almost 60 years ago [RAND, 2001]. We obtained one million random decimal digits from Excel 2003 (Service Pack 2) and Excel 2010 using 200,000 consecutive calls to the RAND() function formatted as a five digit integer with leading zeroes. We created the first million digits of the [Champernowne, 1933] number, CHAMP, which is proven normal. For each expansion we computed BiEn($s$) for the first 1,000 then the first 10,000 and then the first 30,000 consecutive sections each of length $s$ = 32 bits starting at the first digit. From the decimal expansions, we encoded digits 0-4 as 0 and 5-9 as 1. We show the mean and standard deviation of BiEn for each set of strings in Table 5.

Table 5 - The BiEntropy of Some Irrational, Random and Normal Numbers

| Number/Set | Mean N=1,000 | Stdev | Mean N=10,000 | Stdev | Mean N=30,000 | Stdev |
|---|---|---|---|---|---|---|
| PI | 0.6190 | 0.3367 | 0.6302 | 0.3346 | 0.6283 | 0.3361 |
| EXCEL10 | 0.6260 | 0.3422 | 0.6286 | 0.3381 | 0.6281 | 0.3374 |
| CHAMP | 0.6707 | 0.3763 | 0.6180 | 0.3443 | 0.6263 | 0.3353 |
| SQRT(2) | 0.6171 | 0.3400 | 0.6267 | 0.3384 | 0.6254 | 0.3384 |
| EXCEL03 | 0.6271 | 0.3310 | 0.6237 | 0.3364 | 0.6250 | 0.3370 |
| SQRT(3) | 0.6408 | 0.3321 | 0.6307 | 0.3356 | 0.6247 | 0.3375 |
| E | 0.6425 | 0.3341 | 0.6243 | 0.3388 | 0.6235 | 0.3376 |
| RAND | 0.6288 | 0.3382 | 0.6179 | 0.3385 | 0.6217 | 0.3374 |

The BiEntropy of the RAND string is significantly different from (i.e. lower than) the BiEntropy of the decimal expansions of π and the Excel 2010 random digit string ($p < 0.05$) for 10,000 and 30,000 trials and from Champernowne's number for 30,000 trials. The BiEntropies of the irrational, Excel RNG and Champernowne strings are not significantly different from each other for 30,000 trials. Note that there are $2^{32}$ potential entropy states for the 32 bit BiEn function. The sample sizes used in this study constitute a minute proportion of this state space.

We performed a further range of BiEntropic analysis on this data using strings varying in length from 16 bits up to 256 bits. We used BiEn for the shorter string lengths and TBiEn for string lengths > 32. We used in addition an alternative encoding whereby each decimal digit 0-7 generated a bit string 000-111, decimal 8 generated 0 and decimal 9 generated 1. We did not find any statistically significant differences between the BiEntropies of any of these strings save the expected occasional spurious result. Except that we found the result for the RAND string compared with π, Excel 10 and Champernowne's number confirmed ($0.1 < p < 0.001$) in the alternative binary expansion when measured by the 32 bit BiEn function for 75,000 trials.

We show in Figure 5 the frequency distribution of BiEntropy for all eight digit strings for N=1,000 trials. We show in Figure 6 the initial slow progress of convergence by averaging BiEntropy for the first 2, 3, 4 …1000 values for each series. Other work has explored deficits from maximal irregularity for the same irrationals [Pincus & Kalman, 1997].

**Figure 5 Frequency distribution of BiEn for the first 1,000 32-bit strings**

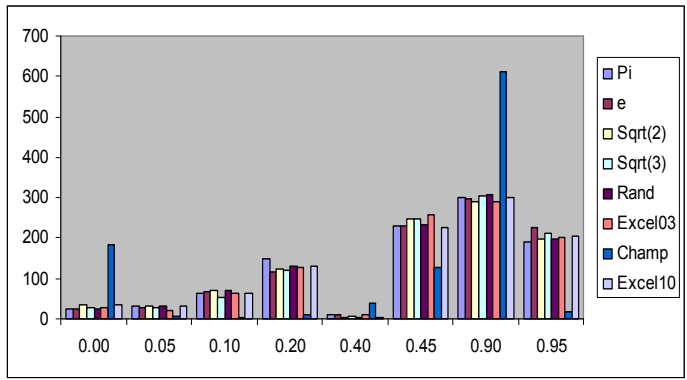

**Figure 6 Mean BiEntropy for the first 1000 32-bit strings**

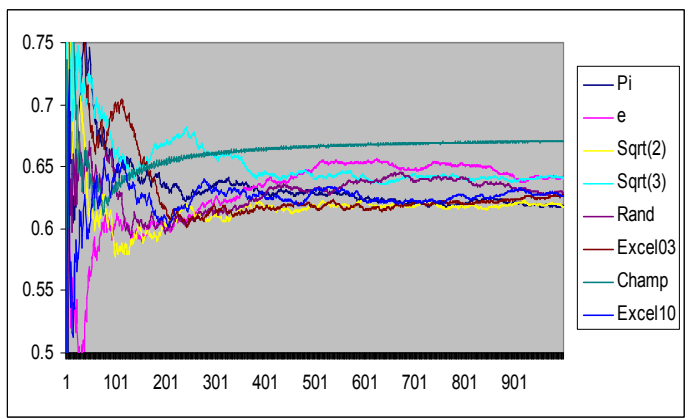

## 12 BIENTROPY OF SOME PLAIN AND ENCRYPTED SPREADSHEETS

We created and/or obtained four large Excel Spreadsheets each of which had a file length of approximately 175 Kilobytes. We encrypted each spreadsheet using the weak Office 97 and strong 128 bit AES algorithms supported in Excel 2003. We computed TBiEn, using a simple C routine, for 1,000 sections each of length 1,024 bits for each encrypted or unencrypted spreadsheet file using the unmodified raw binary data of each file. The contents of each spreadsheet and the values of TBiEn are given in Table 6.

**Table 6 - The Logarithmic BiEntropy of Some Plain and Encrypted Spreadsheets**

| Spreadsheet | Encryption | TBiEn (1024 bit) N=1,000 | |
|---|---|---|---|
| | | **Mean** | **Stdev** |
| Numbers (all cells = 123) | None | 0.8980 | 0.0732 |
| Numbers (all cells = 123) | Office 97 | 0.9913 | 0.0545 |
| Numbers (all cells = 123) | AES | 0.9913 | 0.0545 |
| Random Numbers | None | 0.9857 | 0.0559 |
| Random Numbers | Office 97 | 0.9913 | 0.0545 |
| Random Numbers | AES | 0.9914 | 0.0545 |
| Address Database | None | 0.9428 | 0.1770 |
| Address Database | Office 97 | 0.9913 | 0.0545 |
| Address Database | AES | 0.9913 | 0.0545 |
| Financial Model | None | 0.9450 | 0.1736 |
| Financial Model | Office 97 | 0.9912 | 0.0545 |
| Financial Model | AES | 0.9911 | 0.0545 |

The BiEntropies of all the encrypted files differ from their unencrypted counterparts ($p < 0.01$). All the unencrypted spreadsheets are distinguished from each other ($p < 0.01$) except for the Address Database / Financial model pairing. The BiEntropy reflects the file contents. The unencrypted file with the lowest entropy had the numeric constant 123 in every cell. The unencrypted file with the highest entropy had random number values in every cell. The entropy of the address database and the financial model were similar, lying between the other two extremes. BiEntropy has not distinguished between the two encryption methods. BiEntropy has, of course, no knowledge of the Excel file structure. Figure 2 implies that only half of all binary strings are fully aperiodic, which may have implications in the cryptographic security of key bit strings.

## 13 BIENTROPY OF UK STOCK MARKET PRICE CHANGES

We obtained [Yahoo, 2012] the UK FTSE daily closing prices for the near ten year period 1/1/2003 – 22/8/2012. These were provided in an Excel spreadsheet with the rows representing the days and the columns representing the 100 largest companies in the FTSE index. We deleted rows corresponding to weekends and bank holidays, and columns where the company had not been FTSE quoted for the entire period. The resulting spreadsheet had 2,452 rows $i$ and 71 columns $j$ containing FTSE data in a contiguous array $P_{i,j}$ with no zero values.

We created a second binary array $T_{i,j}$ which recorded absolute price changes between one day and the next, starting at the second day (first day = 0). We used a threshold $R$, ($0.00 \leq R \leq 0.03$) such that:

$$\textit{if } \text{ABS} ( P_{i,j} / P_{i-1,j} - 1 ) > R \textit{ then } T_{i,j} = 1 \textit{ else } T_{i,j} = 0$$

We used the Excel RAND () function and an Excel Data Table [Tyszkiewicz & Balson, 2012] to facilitate the repeated selection of 1000 start prices $S_k$ from $P_{i,j}$ where $i$ and $j$ were uniform random ($1 \leq i \leq 2200$ and $1 \leq j \leq 71$) and a corresponding set of closing prices $C_k$, where $C_k = P_{i+d,j}$ ($32 \leq d \leq 63$).

We calculated a set of holding returns

$$H_k = C_k / S_k$$

And a corresponding set of 32 bit BiEntropies $B_k$ using $T_{i,j}$ through $T_{i+31,j}$.

We implemented the spreadsheet such that we created values of $B_k$ and $H_k$ separately for each of 32 values of $d$ and 7 values of $R$. We sorted the Holding Returns $H_k$ by descending BiEntropy $B_k$ and computed the mean holding return for the upper and lower BiEntropy deciles (by summing the 100 closing prices and dividing by the sum of the 100 opening prices for each of the two deciles). When analysing the Mean Holding Returns and the BiEntropies for each of the 7 individual values of $R$, we were unable to find any statistical significance for $d$. We show in Table 7 the Holding Return observed for all $R \geq 0$ for all $d$ for the upper and lower BiEntropy deciles.

### Table 7 High & Low BiEntropy - Mean Holding Return for various $R$

| $R$ | Upper BiEntropy Decile Mean Holding Return | S.D. | Lower BiEntropy Decile Mean Holding Return | S.D. | $p$ ($n=32$) | Sparsity% $T_{i,j}$ |
|---|---|---|---|---|---|---|
| 0.000 | 1.0249 | 0.0328 | 1.0393 | 0.0406 | - | 49.04 |
| 0.005 | 1.0442 | 0.0401 | 1.0238 | 0.0437 | < 0.10 | 36.55 |
| 0.010 | 1.0552 | 0.0513 | 0.9964 | 0.0196 | < 0.01 | 25.44 |
| 0.015 | 1.0593 | 0.0573 | 1.0000 | 0.0141 | < 0.01 | 17.45 |
| 0.020 | 1.0543 | 0.0443 | 1.0030 | 0.0155 | < 0.01 | 11.95 |
| 0.025 | 1.0452 | 0.0509 | 1.0054 | 0.0115 | < 0.01 | 8.38 |
| 0.030 | 1.0342 | 0.0301 | 1.0098 | 0.0101 | < 0.01 | 6.00 |

We replaced a very small number of outliers where the Holding return was more than 3 S.D. from the mean for $R < 0.03$. For $R = 0.03$ the Holding Return for the lower BiEntropy decile was probably bimodal as 7 closely spaced results with a Holding Return $<< 1.00$ were replaced. The significance and sense of the reported result was unchanged. For $R > 0.02$, the sparsity of $T_{i,j}$ was such that the lower BiEntropy decile contained more than 100 zero entries.

The mean holding return for FTSE stocks held for an average period of 48 days was approximately 4% higher for stocks which had exhibited more disordered (higher BiEntropic) behaviour at the beginning of the observed period compared with stocks which had exhibited more periodic (lower BiEntropic) behaviour ( $p << 0.01$). The effect was not observed without a threshold $R \geq 0.005$.

Subsequent analysis of the data for all $R > 0$ showed that $d$ also had a small positive effect of 1.3% for $d = 32$ on the holding return ($p < 0.05$).

The above analysis was performed using the 32 bit logarithmic TBiEn function. There were no statistically significant results to report using the 32 bit power law version of BiEn.

## 14 SUMMARY

We have described the BiEntropy algorithm and investigated its basic performance on 2, 4 and 8 bit binary strings. We have demonstrated that we can rank binary strings of arbitrary length in terms of the relative order and disorder of all of their digits. Our method is very simple and is based upon the use of the Exclusive Or function and some arithmetic weights. We show that BiEntropy is fractal such that the ordering method is consistent across strings of arbitrary length.

We used the BiEntropy function to investigate the order and disorder in the sequence of prime numbers. We showed that BiEntropy determined that the sequence of early prime numbers was disordered. We proved explicitly that the sequence of prime numbers is not periodic.

We then evaluated some 5*7 ISO dot matrix printer characters and some randomly generated characters on a similar 5*7 grid. Using the insight gained from BiEn, we designed a Low Entropy Character Set (LECS) and compared these characters with others from a High Entropy Character Set (HECS) which we also designed. We measured the BiEntropy of the characters of the Braille character set and showed for the first time that Braille is also a low entropy character set. The LECS character set is visually distinctive and easily extensible. It may provide a rational basis for extending the Braille methodology to a significantly wider character set and user base.

We then used BiEntropy to determine the relative order and disorder of consecutive 32 digit sections of some long expansions of some well-known irrational numbers including *pi* and *e* which we compared with Champernowne's normal number and outputs from the Excel 2003 and Excel 2010 RNG's. Despite the well known problems [McCullough, 2008] with the Excel RNG's we show that we cannot distinguish between the Excel RNG output and normal and suspected normal numbers using the 32 bit BiEntropy method.

We examined some plain and encrypted Excel spreadsheets and show that BiEntropy is able to clearly distinguish between them in the absence of any knowledge of the Excel file structure.

We examined ten years worth of daily UK stock market prices. We showed that investors prefer – i.e. pay more for – stocks that exhibit prior disordered behaviour. This result, if confirmed, will have significant implications for global financial markets [Croll, 2007 & 2009]. It is possible that entropy options may emerge as a means of profiting from and then neutralising the illustrated effects. There is in principle hidden information in any time series. Statistical time series methods may need to be re-examined in the light of these results.

## 15 FURTHER WORK

Given the successful application of BiEntropy in the five diverse fields outlined above, it is likely that BiEntropy will be widely applicable. We are working on the use of BiEntropy in Bit String Physics [Noyes, 1997]. We keenly anticipate any results which may follow from BiEntropic analysis of the binary encoding of the four bases of the human genome.

## ACKNOWLEDGEMENTS

The author thanks: Lars and James Mansson for obtaining and formatting the FTSE data; ANPA 34 colleagues for affirming Corollary One; Tony Bartlett for patiently listening to the first exposition of Corollary Two; Qinqin Zhang, Math librarian at the Library of the University of Western Ontario for locating and scanning a copy of James Mc Nair's MSc Thesis report; Dr Mike Manthey and Dr John C. Amson for their reviews of various drafts; My wife Deborah for her patience during my effective absence during the early part of this work in Jan & Feb 2013. This work was financed by the author during the period December 2012-Aug 2013. There are no conflicts of interest to declare. All data is naturally or will be made publicly available. This paper is dedicated to the memory of Louis Braille (1809-1852) - *ça va bien et merci mon ami*.